\documentclass{PoS}
\usepackage{amsmath,amssymb}  
\usepackage{braket}

\newcommand{\Tprod}[1]{{\mathrm T}\lbrack #1 \rbrack}
\allowdisplaybreaks

\title{Electroweak and non-resonant corrections to top-pair production near threshold at NNLO}

\ShortTitle{Electroweak and non-resonant corrections to top-pair production near threshold at NNLO}

\author{Martin Beneke\\
        Physik Department T31, James-Franck-Stra\ss{}e~1, Technische Universit\"at M\"unchen, \\
        85748 Garching, Germany}

\author{Andreas Maier\\
        IPPP, Department of Physics, University of Durham, DH1 3LE, United Kingdom}
        
\author{\speaker{Thomas Rauh}\thanks{TUM-HEP-1105/17, IPPP/17/81, FTUAM-17-25, IFT-UAM/CSIC-17-106}\\
        IPPP, Department of Physics, University of Durham, DH1 3LE, United Kingdom\\
        E-mail: \email{thomas.j.rauh@durham.ac.uk}}

\author{Pedro Ruiz-Femen\'ia\\
        Departamento de F\'isica Te\'orica and Instituto de F\'isica Te\'orica UAM-CSIC, \\
        Universidad Aut\'onoma de Madrid, E-28049 Madrid, Spain}

\abstract{The top-quark mass can be determined with very high precision from a scan 
of the total $e^+e^-\to b\bar{b}W^+W^-X$ cross section near the top-pair production 
threshold. 
We present the full calculation of electroweak and non-resonant corrections to this 
process at NNLO. We discuss the size of the new contributions and estimate the theory 
uncertainty on the top-quark mass in the PS mass scheme. 
}

\FullConference{13th International Symposium on Radiative Corrections (Applications of Quantum Field Theory to Phenomenology)\\
                 25-29 September, 2017 \\
                 St. Gilgen, Austria}

\begin{document}


\section{Introduction\label{sec:intro}}

The total inclusive cross section for the process $e^+e^-\to b\bar{b}W^+W^-X$ 
is highly sensitive to the value of the top-quark mass in the vicinity of the 
top-pair production threshold $\sqrt{s} = 2m_t$. At a future lepton collider 
this facilitates the determination of the top-quark mass in a well-defined 
mass scheme like $\overline{\text{MS}}$ or PS~\cite{Beneke:1998rk} with an 
uncertainty of about 50 MeV~\cite{Simon:2016pwp,Beneke:2015kwa}. 
In addition the strong coupling constant and the top-quark width and Yukawa 
coupling can be determined. Given that theoretical uncertainties dominate 
over statistical ones, precise predictions for the cross section are crucial 
for this programme. 

The challenge in higher-order computations in the threshold region lies in 
dealing with the non-relativistic nature of the process. The non-relativistic 
top pair interacts strongly through a non-local color Coulomb potential which 
manifests itself at the level of the cross section as corrections scaling with 
powers of $\alpha_s/v$, where $v$ is the top-quark velocity. These effects 
become non-perturbatively strong in the threshold region, where $v\sim\alpha_s$, 
and must be resummed. The techniques to perform this resummation are based on 
effective field theories (EFT) and exploit the hierarchy between the dynamical 
scales $m_t$ (hard), $m_t v$ (soft) and $m_t v^2$ (ultrasoft) of the process. 
By integrating out the hard and soft modes the problem of solving multiscale 
Feynman integrals is divided into simpler calculations of matching coefficients 
and of a non-relativistic Green function. Within this approach the QCD 
corrections have been computed up to NNNLO~\cite{Beneke:2015kwa}, where the 
EFT framework is given by potential non-relativistic QCD 
(PNRQCD)~\cite{Pineda:1997bj,Beneke:2013jia}. 

\begin{figure}
 \centering
 \includegraphics[width=0.4\textwidth]{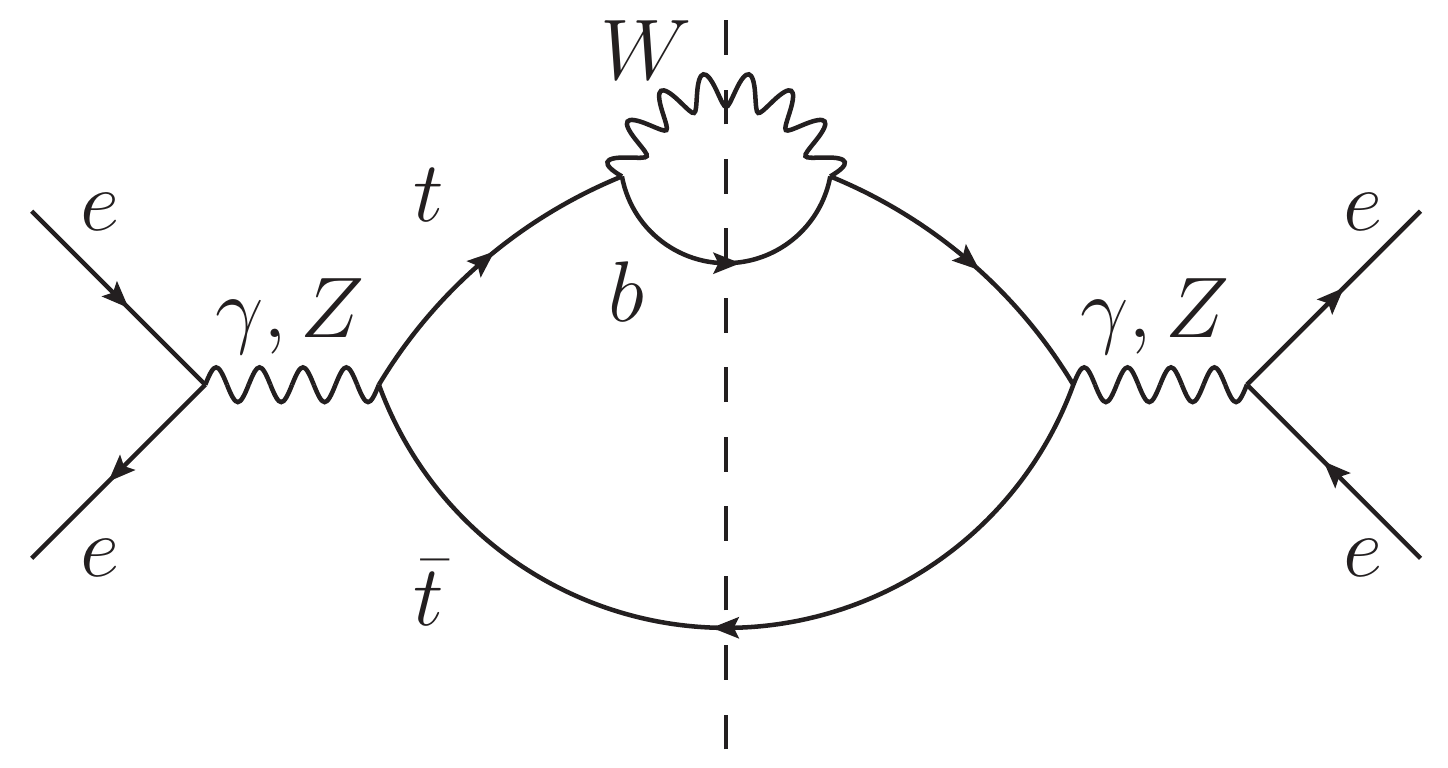}\hspace{0.8cm}
 \includegraphics[width=0.4\textwidth]{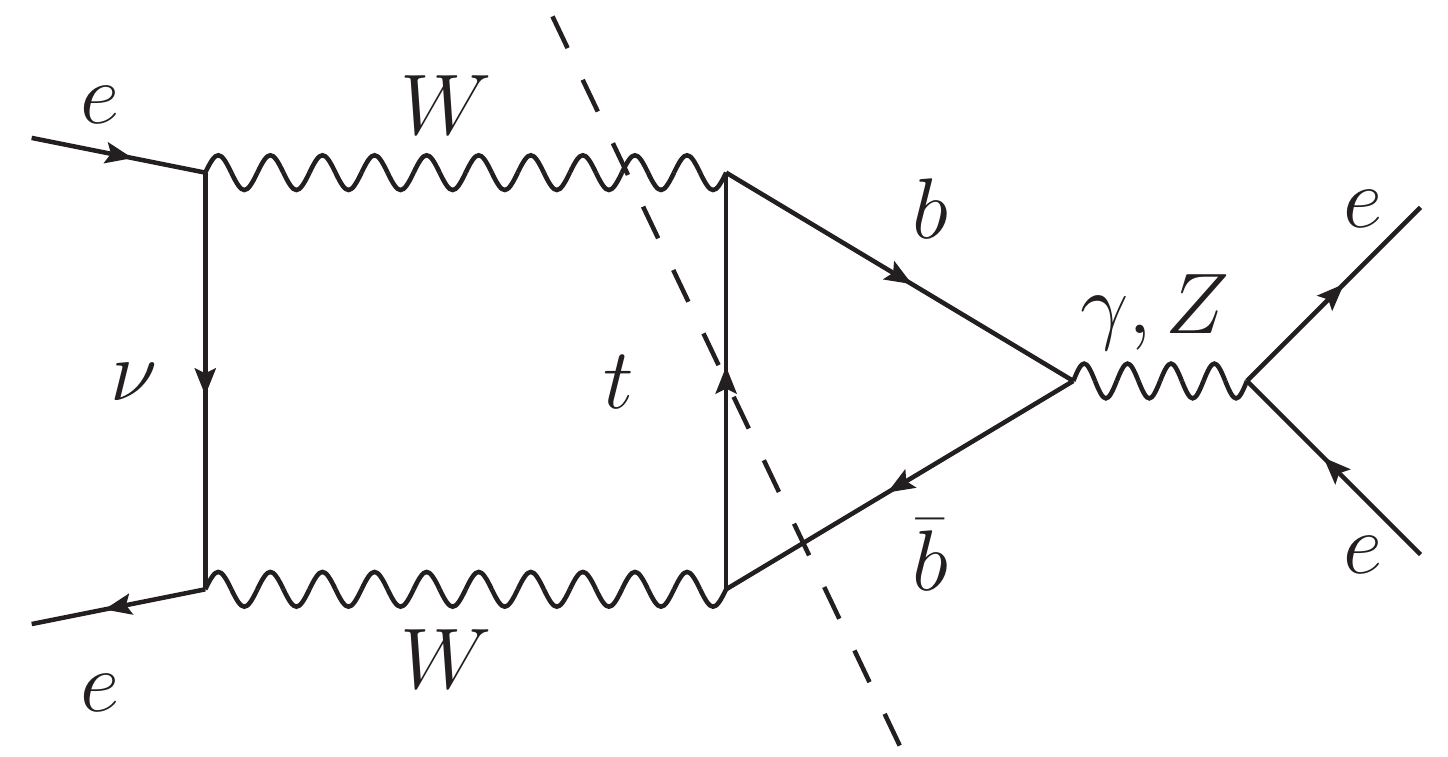}
 \caption{Example diagrams contributing to the non-resonant cross section 
 at NLO. In the left diagram, the top lines are off-shell, and the antitop 
 ones near mass-shell.\label{fig:nonres_example}}
\end{figure}

The scale uncertainty of the NNNLO QCD result is at the level of only $\pm3\%$. 
We find, however, that non-QCD effects at NLO yield corrections up to 
15\%~\cite{Beneke:2010mp,Beneke:2015lwa}, which motivates the calculation of the 
full NNLO non-QCD corrections~\cite{Beneke:2017NNLO} presented in these proceedings. 
The extension of the calculation beyond pure QCD requires a systematic 
treatment of the instability of the top quark. 
Counting $\alpha_{\rm EW}\sim\alpha_s^2\sim v^2$, the top-quark decay width 
$\Gamma_t\sim m_t\alpha_{\rm EW}$ is of the same order as the ultrasoft scale $m_t v^2$, 
which implies that the narrow width approximation for the top quarks is 
unphysical and that we \emph{must} consider the final state $b\bar{b}W^+W^-X$ 
after top decay. Consequently, we have to take into account not only the resonant 
production of the final state through the decay of an intermediate non-relativistic 
top pair, but also the non-resonant production through hard processes as shown in 
Figure~\ref{fig:nonres_example}. 
The example diagrams are of the relative order $\alpha_{\rm EW}/v\sim\alpha_s$, where 
the $1/v$ factor accounts for the phase-space suppression for the production 
of a non-relativistic particle pair in the resonant part which is not present 
in the non-resonant contribution, and therefore contribute at NLO. 

For the systematic calculation of the non-QCD effects the EFT framework must be 
extended to Unstable Particle Effective Theory~\cite{Beneke:2003xh,Beneke:2004km}. 
Utilizing the optical theorem, the cross section takes the form 
\begin{align}
 \sigma(s) \sim \text{Im}\Bigg[ & \, \sum_{k,l} \, C^{(k)} C^{(l)} \int d^4 x \,
\braket{e^- e^+ |\Tprod{i {\cal O}^{(k)\dagger}(0)\,i{\cal O}^{(l)}(x)}|e^- e^+} \notag\\
 & + \sum_{k} \, C_{4 e}^{(k)} \braket{e^- e^+|i {\cal O}_{4e}^{(k)}(0)|e^- e^+}\Bigg],
 \label{eq:master_formula}
\end{align}
where only cuts corresponding to the final state $b\bar{b}WWX$ must be considered 
when taking the imaginary part. The first line of~\eqref{eq:master_formula} 
corresponds to the resonant cross section, where the production operator 
${\cal O}^{(l)}$ annihilates the external $e^+e^-$ state and produces a 
non-relativistic top pair. The hard matching coefficients $C^{(l)}$ absorb 
the corrections to the production from hard modes. 
The matrix element must be evaluated with the EFT given by the Lagrangian 
\begin{equation}
 \mathcal{L} = \mathcal{L}_\text{PNREFT} + \mathcal{L}_\text{SCET}^{(-)} + \mathcal{L}_\text{SCET}^{(+)},
\end{equation}
where $\mathcal{L}_\text{PNREFT}$ follows from the generalization of PNRQCD 
to electroweak effects and contains the interactions of the non-relativistic 
top quarks through potentials and with ultrasoft modes and 
$\mathcal{L}_\text{SCET}^{(\mp)}$ contains the interactions of the collinear 
electron and positron with collinear and ultrasoft modes.\footnote{It is also 
possible to integrate out generic collinear modes, keeping only external-collinear 
modes whose momentum differs from that of the external electron or positron 
by an ultrasoft amount, see~\cite{Beneke:2003xh,Beneke:2004km}.} There are 
no interactions of the top quarks with collinear modes in the EFT, because 
momentum conservation implies that they create hard modes which have been 
integrated out. 

The second line of~\eqref{eq:master_formula} corresponds to the non-resonant 
contribution, which is given by the matrix elements of local four-electron 
operators since the non-resonant production is a hard process and the hard 
scale has been integrated out in the EFT. The hard matching coefficients 
$C_{4 e}^{(k)}$ contain imaginary parts from cuts corresponding to the 
$b\bar{b}W^+W^-X$ final state. In practice, we do not determine the coefficients, 
but compute the non-resonant cross section directly. The full NLO contribution 
is given by the sum of the cross sections for the processes 
$e^+e^- \to \bar{t}b W^+$ and $e^+e^- \to t\bar{b} W^-$ evaluated exactly at 
the threshold $s=4m_t^2$ and has been computed in~\cite{Beneke:2010mp}. 
Setting $s=4m_t^2$ isolates the contribution from the hard momentum region, 
since all other modes only yield scaleless integrals that vanish in dimensional 
regularization. The NNLO corrections are given by the $\mathcal{O}(\alpha_s)$ 
corrections to these processes. Doubly non-resonant contributions scale as 
$\alpha_{\rm EW}^2/v\sim\alpha_s^3$ relative to the LO and need not be considered at NNLO. 

The resonant and non-resonant contributions contain finite-width and endpoint 
divergences, respectively, which cancel in the sum~\eqref{eq:master_formula} 
over both parts. Such spurious divergences are common when momentum regions 
are separated~\cite{Beneke:1997zp,Jantzen:2011nz}. In the resonant part, 
finite-width divergences appear in the UV region of the loop integration 
over the momenta of the non-relativistic top quarks, i.e. in the limit where 
top quarks that are parametrically close to resonance are far off-shell. 
They have been regularized dimensionally and we must employ the same scheme 
for the computation of the non-resonant part. The origin of the divergences 
in the non-resonant part is discussed below.


\section{Non-resonant part at NNLO\label{sec:nonresonant}}

As discussed above the non-resonant contribution up the NNLO is given by the 
sum of the processes $e^+e^- \to \bar{t}bW^+$ and $e^+e^- \to t\bar{b}W^-$ 
at threshold at NLO in QCD. This is not a standard $2\to3$ NLO calculation 
because of the presence of endpoint divergences which must be regularized 
dimensionally. To illustrate this issue we consider the phase-space integral 
of one of the tree level diagrams
\begin{equation}
\begin{aligned}
 & \int d\text{LIPS}_{e^+e^-\rightarrow\bar{t}bW^+}\,f_{i}(p_{e^+},p_{e^-},p_{\bar{t}},p_{W^+},p_b)
 \\
 = &\,\, \frac{m_t^2}{2\pi}\int\limits_x^1dt \int d\text{LIPS}_{e^+e^-\rightarrow t^*\bar{t}} \int d\text{LIPS}_{t^*\rightarrow bW^+}\,f_{i}(p_{e^+},p_{e^-},p_{\bar{t}},p_{W^+},p_b)\\
 \equiv &\, \int_x^1dtg_{i}(t), 
\end{aligned}
\label{eq:PSI}
\end{equation}
where standard phase-space integration techniques have been applied to 
split the $2\to3$ process into $2\to1+1^*$ and $1^*\to2$ parts with 
an additional integration over the invariant mass $t=p_{t^*}^2/m_t^2$ 
of the off-shell top quark. The integration limits $x=m_W^2/m_t^2$ and 
1 follow from the kinematic restrictions when the bottom-quark mass is 
neglected. The divergences originate from the endpoint $t\to1$ of the 
integration, where the top-quark, which is parametrically off-shell, 
becomes resonant. This is exactly the opposite of the scenario in which 
the finite-width divergences occur in the resonant part. E.g. the integrand 
of the left diagram in Figure~\ref{fig:nonres_example} takes the form 
\begin{equation}
 g_{h_1}(t) \;\; \mathop{\propto}\limits^{t\to1} \;\; \frac{(1-t)^{1/2-\epsilon}}{(1-t)^2}, 
 \label{eq:gh1}
\end{equation}
where the numerator is a phase-space suppression factor and the denominator 
comes from the two top-quark propagators. We note that the width-dependent 
term in the denominator $(p_{t^*}^2-m_t^2+im_t\Gamma_t)$ of the top propagator 
has been expanded out in~\eqref{eq:gh1} because in the hard momentum region 
$p_{t^*}^2-m_t^2\sim m_t^2$ is parametrically much larger than 
$m_t\Gamma_t\sim m_t^2\alpha_{\rm EW}$. 
The contribution from this diagram is divergent but finite in dimensional 
regularization, because the endpoint divergence as $t\to 1$ is non-logarithmic. 
The other tree-level diagrams contain at most one top-quark propagator and are 
finite~\cite{Beneke:2010mp}, because there are only integrable divergences as 
$t\to 1$. 

At NNLO this holds no longer true and explicit $1/\epsilon$ poles appear. We 
cannot expand the integrands $g_i$ of the NNLO diagrams in $\epsilon$ without 
spoiling the dimensional regularization of the remaining $t$-integration. 
However, the computation of the full $\epsilon$-dependence of the $g_i$, which 
contain phase-space and (for the virtual corrections) loop integrations, 
is difficult. We therefore use subtractions for the endpoint divergences 
\begin{equation}
\int_x^1dt\,g_{i}(t)=\int\limits_x^1dt\left[g_{i}(t)-
\!\sum_{a=1,\frac32,2}\;\sum_{b}\;\frac{\hat{g}_{i}^{(a,b)}}{(1-t)^{a+b\epsilon}}
\right]+
\!\sum_{a=1,\frac32,2}\;\sum_{b}\;\frac{\hat{g}_{i}^{(a,b)}(1-x)^{1-a-b\epsilon}}
{1-a-b\epsilon},
\label{eq:EPsubtraction}
\end{equation}
where the expression in square brackets is finite and can be expanded in 
$\epsilon$ before the integration over $t$ and the second term is the 
integrated subtraction term. The subtraction terms are given by the singular 
terms in $(1-t)$ with the full $\epsilon$ dependence and have been determined 
in~\cite{Jantzen:2013gpa,Beneke:2017NNLO} using the expansion by 
regions~\cite{Beneke:1997zp,Jantzen:2011nz}. The NNLO contributions 
contain endpoint singular terms scaling as $(1-t)^{-(a+b\epsilon)}$ 
with $a=2,3/2,1$. Only those with $a=1$ yield explicit poles in $\epsilon$, 
whereas the other terms are divergent but finite in dimensional regularization. 

We compute the contributions from the endpoint divergent diagrams manually 
using the subtractions~\eqref{eq:EPsubtraction}. For the endpoint finite 
diagrams we use \texttt{MadGraph}~\cite{Alwall:2014hca} code from which we 
remove the contributions from the endpoint divergent diagrams. 
The details of this computation are given in~\cite{Beneke:2017NNLO}.


\section{Electroweak corrections to the resonant part at NNLO\label{sec:resonant}}

The resonant part receives corrections due to the decays of the top quarks and 
other electroweak contributions. At NNLO, we have to consider $\mathcal{O}(\alpha_{\rm EW})$ 
corrections to the hard matching coefficients $C^{(k)}$~\cite{Hoang:2006pd}, 
which become complex because of $\bar{t}bW^+$ and $t\bar{b}W^-$ 
cuts~\cite{Hoang:2004tg,Beneke:2017NNLO}. Furthermore, there are corrections 
to the matrix element from the QED Coulomb potential $-e_t^2\alpha_{\rm em}/r$ and 
higher-order terms involving the top-quark width, e.g. time dilatation effects. 
Last but not least, we have to consider photon radiation from the initial 
state (ISR). The ISR yields large logarithms $\ln(m_e^2/s)$. Thus, in addition 
to the fixed order $\mathcal{O}(\alpha_{\rm EW})$ ISR corrections required at NNLO, 
we resum logarithms at the LL level using the structure function 
approach where the cross section takes the form 
\begin{equation}
 \sigma_\text{w. ISR}(s)=\int\limits_0^1dx_1\int\limits_0^1dx_2 \Gamma_{ee}^\text{LL}(x_1)\Gamma_{ee}^\text{LL}(x_2)\sigma(x_1x_2s).
 \label{eq:convolutionstructurefunctions}
\end{equation}
The structure functions $\Gamma_{ee}^\text{LL}(x)$ give the probability of 
finding an electron with momentum $x\,p$ in the `parent electron' with 
momentum $p$ and can be found in~\cite{Beenakker:1996kt}. They are currently 
not known at NLL. A detailed discussion of the corrections to the resonant 
part is given in~\cite{Beneke:2017NNLO}.


\section{Phenomenology\label{sec:pheno}}

\begin{figure}
 \centering
 \includegraphics[width=0.6\textwidth]{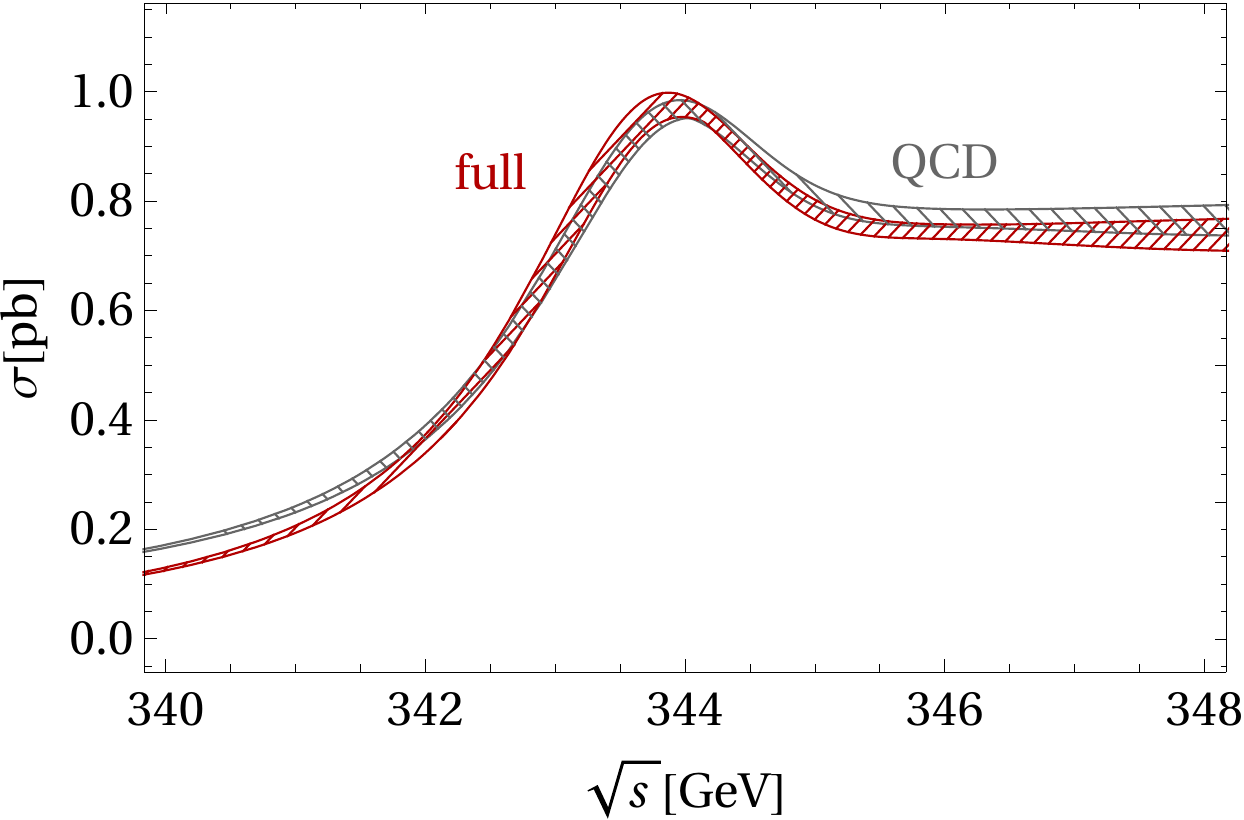}\\[0.5cm]
 \includegraphics[width=0.6\textwidth]{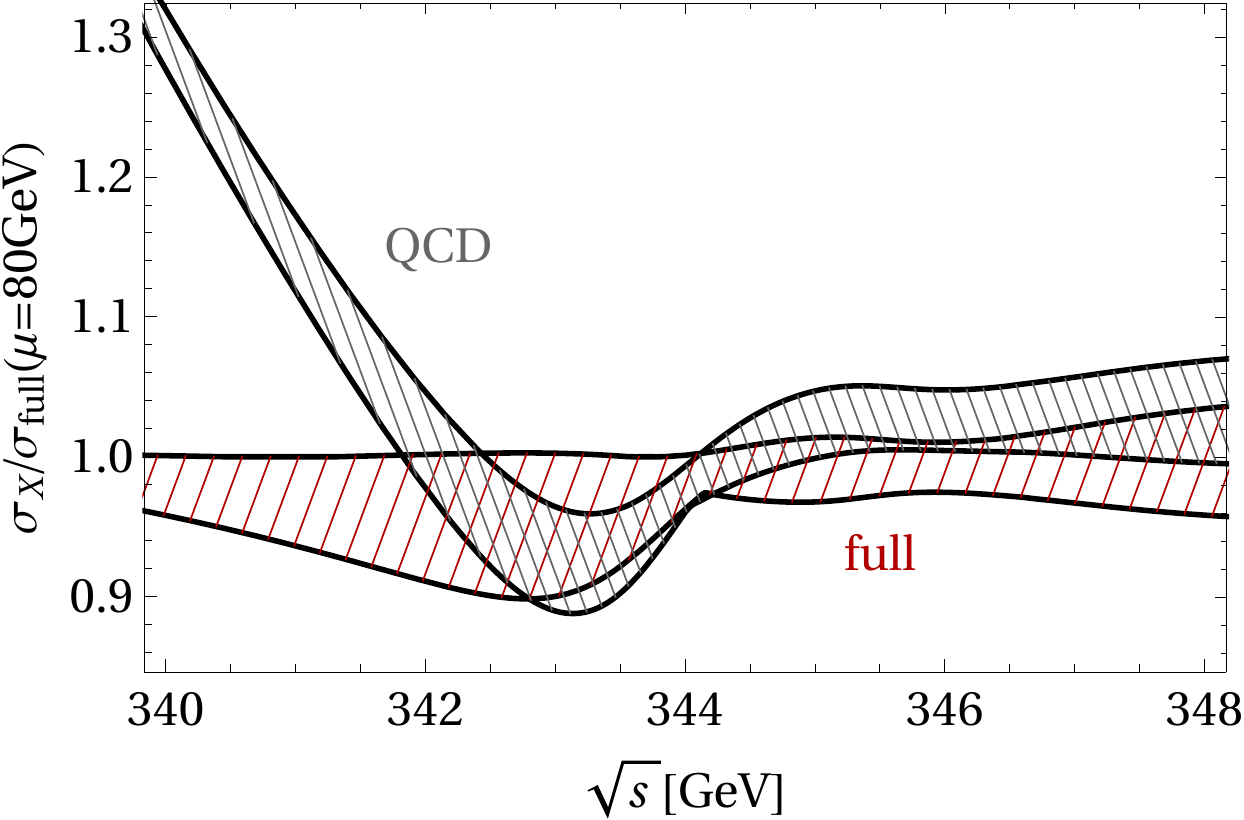}
 \caption{The pure QCD (gray) and full (red) prediction for the total cross 
 section near threshold including the uncertainty from variation of the 
 renormalization scale between 50 and 350 GeV. Shown are the cross section 
 in pb (top) and normalized to the full cross section for the central 
 scale $\mu_r=80\,$GeV (bottom).\label{fig:SigmaAll}}
\end{figure}

The new results have been added to a new version of the \texttt{QQbar\_Threshold} 
code~\cite{Beneke:2016kkb}, which will soon become public. The effect of the 
non-QCD corrections is shown in Figure~\ref{fig:SigmaAll}. The input values are 
$m_t^\text{PS} = 171.5\,\text{GeV}$, $\Gamma_t = 1.33\,\text{GeV}$ and the default 
values of the code for all other parameters. Our central scale choices are 
$\mu_r = 80\,\text{GeV}$ and $\mu_w = 350\,\text{GeV}$ where $\mu_r$ is the 
renormalization scale and $\mu_w$ is the scale associated with the finite-width 
divergences. The dependence on the latter cancels exactly at NNLO when all 
corrections are included and only a tiny dependence remains at NNNLO where not 
all contributions are known. 
The QCD cross section is the sum of the S-wave~\cite{Beneke:2015kwa} and 
P-wave~\cite{Beneke:2013kia} contributions while the full results includes 
in addition all NNLO corrections~\cite{Beneke:2017NNLO} and Higgs effects 
at NNNLO~\cite{Beneke:2015lwa}, but no ISR effects. The bands in 
Figure~\ref{fig:SigmaAll} are spanned by variation of the renormalization 
scale between 50 and 350\,GeV. The largest corrections are observed below 
the threshold where the cross section is reduced by up to 25\%. This is 
where the cross section itself becomes small and the almost energy-independent 
non-resonant contribution yields a large relative correction. The non-QCD 
corrections also modify the shape of the cross section, making the remnant 
of the 1S toponium peak more pronounced. 

\begin{figure}
 \centering
 \includegraphics[width=0.6\textwidth]{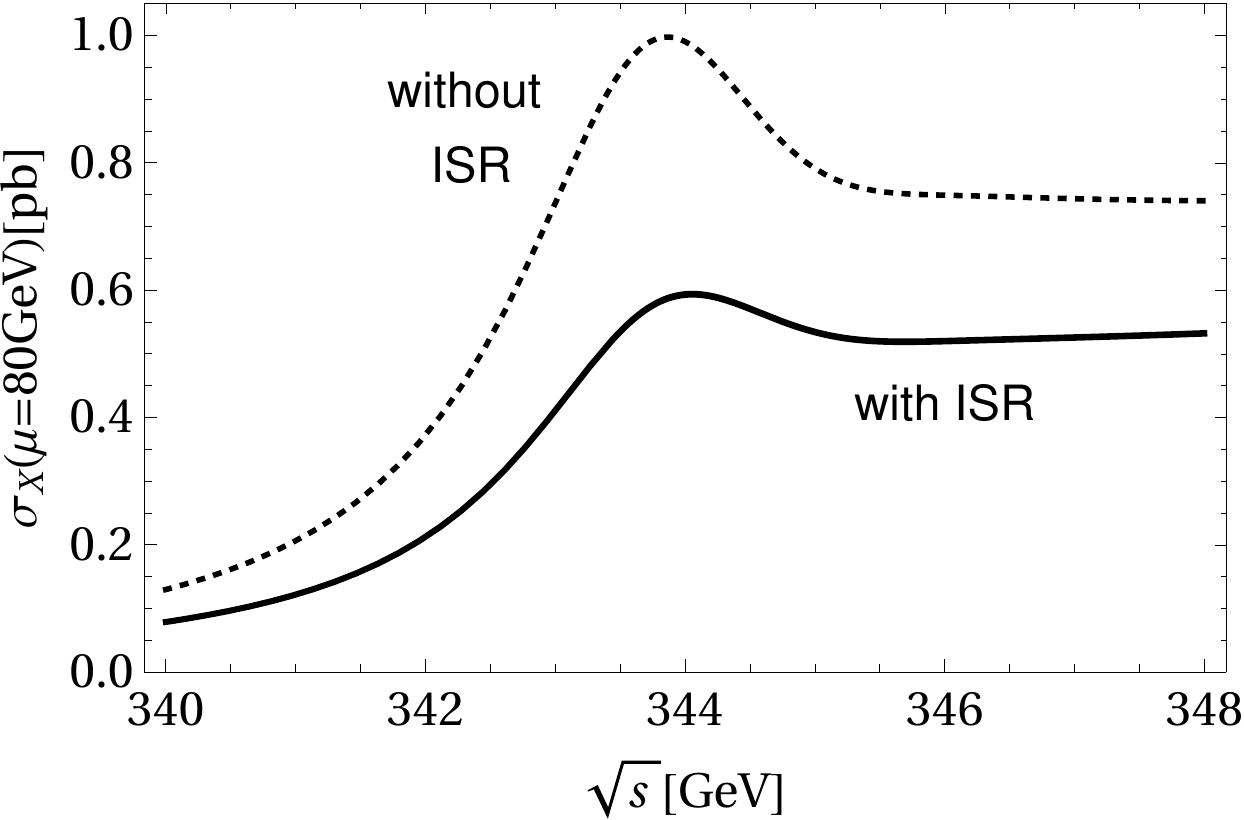}
 \caption{The effect of initial state radiation on the cross section. \label{fig:ISR}}
\end{figure}

The ISR corrections are shown in Figure~\ref{fig:ISR} and reduce the cross 
section by 30 - 45\,\%. The peak is smeared out and shifted to the right 
by almost 200\,MeV. This emphasizes the need for a full NLL description of 
ISR which is crucial for precision physics at a future lepton collider. 
We also note that when electromagnetic initial-state corrections are accounted 
for at NNLO, the definition of the ISR convolution is factorization-scheme 
dependent and one can no longer use a phenomenological convolution as is often 
done in experimental studies without reference to the factorization scheme 
defining the cross section without ISR. 

\begin{figure}
 \centering
 \includegraphics[width=0.6\textwidth]{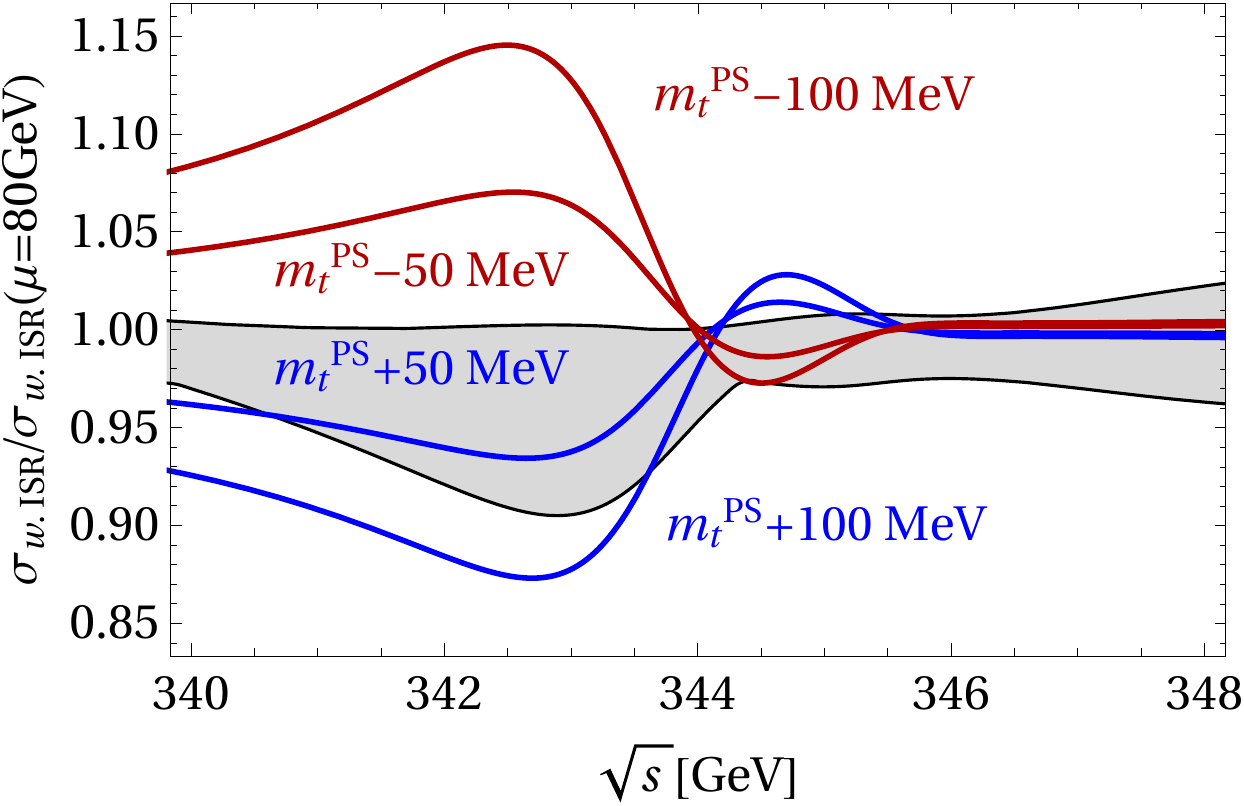}
 \caption{The effect of a variation of the input value of the top-quark mass 
 on the prediction for the cross section. The scale uncertainty band of the 
 full cross section is shown for comparison. The predictions include ISR effects 
 with the accuracy described in Section~\ref{sec:resonant}. 
 \label{fig:mass}}
\end{figure}

The sensitivity of the cross section to the top-quark mass can be estimated 
by comparing the effects of a variation in the input value to the theoretical 
uncertainty as shown in Figure~\ref{fig:mass}. To a good approximation a 
shift $\delta m_t$ in the value of the top-quark mass moves the shape of the 
cross section to the right by $2\delta m_t$. The region where the slope is 
large is most sensitive to this shift. Comparing the effect to the scale 
uncertainty we expect that the theoretical uncertainty is of the order of 
$\pm 40$\,MeV. This estimate is in agreement with the results of a more 
realistic simulation~\cite{Simon:2016pwp}, which takes into account the 
luminosity spectrum and the properties of the detector. Including statistical 
uncertainties and experimental systematics, the total uncertainty is expected 
to be about $\pm 50$\,MeV.


\section{Conclusions\label{sec:conclusions}}

We presented results of a computation of the complete NNLO 
non-resonant and electroweak corrections to the total inclusive 
$b\bar{b}W^+W^-X$ production cross section in $e^+e^-$ collisions 
near the top-quark pair production threshold~\cite{Beneke:2017NNLO}. 
The contributions are important both conceptually, since they are 
required to cancel left-over divergences in the pure QCD cross section, 
as well as phenomenologically, since they lead to sizeable modifications 
of the cross section. 
Together with the NNNLO QCD~\cite{Beneke:2015kwa,Beneke:2013kia} 
and Higgs~\cite{Beneke:2015lwa} contributions, the high precision of 
the theory predictions allows a determination of the top-quark mass 
in a well-defined scheme with an uncertainty of about 50\,MeV at a 
future lepton collider. 

Last but not least, we note that the EFT techniques developed for these 
calculations can also be used in other contexts. For instance, the 
understanding of the behaviour of the $gg\to HH$ amplitude near the 
top-pair production threshold has recently been used~\cite{Grober:2017uho} 
to reconstruct the top-quark mass dependence of the two-loop 
amplitude~\cite{Borowka:2016ehy,Borowka:2016ypz} with high accuracy and 
the approach used there can also be applied at higher orders and for 
similar processes.

\end{document}